# Polynomial functions for direct calculation of the surface free energy developed from the Neumann Equation of State method

Jonathan M. Schuster[a,b,c], Carlos E. Schvezov[a], Mario R. Rosenberger[a,b]

[a]*Instituto de Materiales de Misiones (IMAM), Universidad Nacional de Misiones (UNaM), Consejo Nacional de Investigaciones Científicas y Técnicas (CONICET), Félix de Azara 1552, C.P. 3300, Posadas – Misiones – Argentina.*

[b]*Universidad Nacional de Misiones (UNaM), Facultad de Ciencias Exactas, Químicas y Naturales (FCEQyN), Programa de Materiales, Modelización y Metrología, Ruta 12, km 7.5, C.P. 3300, Miguel Lanús, Posadas – Misiones – Argentina.*

[c]*Universidad Nacional del Alto Uruguay (UNAU), Departamento de Ciencias de la Salud, Cátedra de Física Biomédica, Avenida Tejeda 1042, C.P. 3364, San Vicente - Misiones – Argentina.*

*Corresponding author: jschuster@fceqyn.unam.edu.ar

**Abstract**

The Neumann Equation of State (EQS) allows obtaining the value of the surface free energy of a solid ($\gamma_{SV}$) from the contact angle ($\theta$) of a probe liquid with known surface tension ($\gamma_{LV}$). The value of $\gamma_{SV}$ is obtained by numerical methods solving the corresponding EQS. In this work, we analyzed the discrepancies between the values of $\gamma_{SV}$ obtained using the three versions of the EQS reported in the literature. The condition number of the different EQS was used to analyze their sensitivity to the uncertainty in the $\theta$ values. Polynomials fit to one of these versions of EQS are proposed to obtain values of $\gamma_{SV}$ directly from contact angles ($\gamma_{SV}(\theta)$) of particular probe liquids. Finally, a general adjusted polynomial is presented to obtain the values of $\gamma_{SV}$ not restricted to a particular probe liquid ($\gamma_{SV}(\theta, \gamma_{LV})$). Results showed that the three versions of EQS present non-negligible discrepancies, especially at high values of

$\theta$. The sensitivity of the EQS to the uncertainty in the values of $\theta$ is very similar in the three versions and depends on the probe liquid used (greater sensitivity at higher $\gamma_{LV}$) and on the value of $\gamma_{SV}$ of the solid (greater sensitivity at lower $\gamma_{SV}$). The discrepancy of the values obtained by numerical resolution of both the fifth-order fit polynomials and the general fit polynomial was low, no larger than ±0.40 mJ/m². The polynomials obtained allow the analysis and propagation of the uncertainty of the input variables in the determination of $\gamma_{SV}$ in a simple and fast way.

## 1. Introduction

The contact angle of a liquid deposited on a solid surface is known as the most practical and convenient method to determine the surface tension of solids ($\gamma_S$) [1]. One of the most used methods to determine the contact angle ($\theta$) is the sessile drop method, which consists in measuring the contact angle by means of a goniometer of a liquid drop deposited on the surface of the solid of interest [2–4]. The surface tension is then calculated using the Young equation [5–7]:

$$\gamma_{LV} \cos \theta = \gamma_{SV} - \gamma_{SL} \qquad (1)$$

where $\gamma_{SV}$, $\gamma_{SL}$ and $\gamma_{LV}$ are interfacial tensions between solid and vapor, solid and liquid, and liquid and vapor, respectively, as shown in Figure 1. If an ideal solid surface is considered (rigid, perfectly smooth and chemically homogeneous) the value of $\theta$ is unique and is called equilibrium contact angle [8].

In addition, considering that the reduction of the surface tension due to vapor adsorption is negligible, that is, the diffusion pressure $\pi_e = \gamma_S - \gamma_{SV}$ is negligible, and therefore $\gamma_{SV} \cong \gamma_S$ [9], the surface tension of the solid $\gamma_S$ is approximately $\gamma_{SV}$. Thus, assuming this approximation as valid, it allows experimentally determining $\gamma_{SV}$ instead of $\gamma_S$.

One of the most used methods to determine $\gamma_{SV}$ from contact angle measurement is based on the equation of state proposed by Neumann [8–11]. The Neumann method is based on the empirical assumption that the $\gamma_{SV}$ of a determined material is constant, independently of the liquid in contact. Therefore, in the Young equation $\gamma_{LV} \cos\theta$ is only related to $\gamma_{LV}$. In addition, each solid surface produces a unique experimental curve relating $\gamma_{LV} \cos\theta$ vs $\gamma_{LV}$, as shown in Figure 2. These empirical results allow concluding that $\gamma_{LV} \cos\theta$ depends only on $\gamma_{LV}$ and $\gamma_{SV}$ [12,13], as follows:

$$\gamma_{LV} \cos\theta = F_1(\gamma_{LV}, \gamma_{SV}) \tag{2}$$

where $F_1$ is a function to be determined. Combining equations 1 and 2, we obtain:

$$\gamma_{SL} = \gamma_{SV} - F_1(\gamma_{LV}, \gamma_{SV}) = F_2(\gamma_{LV}, \gamma_{SV}) \tag{3}$$

which is called the equation state relation [9,12]. For this relation, Neumann et al. [14] developed an empirically adjusted relation based on contact angle measurement on different solid surfaces, as follows:

$$\gamma_{SL} = \frac{\left(\sqrt{\gamma_{LV}} - \sqrt{\gamma_{SV}}\right)^2}{1 - 0.015\sqrt{\gamma_{LV}\gamma_{SV}}} \tag{4}$$

Combining equation 4 with equation 1, we obtain what is called the first Neumann equation of state model [14] which is a non-explicit function for $\gamma_{SV}$:

$$\frac{(0.015\,\gamma_{SV} - 2.00)\sqrt{\gamma_{LV}\gamma_{SV}} + \gamma_{LV}}{\gamma_{LV}(0.015\sqrt{\gamma_{LV}\gamma_{SV}} - 1)} - \cos\theta = 0 \tag{5}$$

By using a theoretical approach, Li and Neumann [15] obtained the following equation:

$$\gamma_{SL} = \gamma_{LV} + \gamma_{SV} - 2\sqrt{\gamma_{LV}\gamma_{SV}}\, e^{-\beta_1(\gamma_{LV} - \gamma_{SV})^2} \tag{6}$$

which, combined with equation 1, results in the second Neumann equation of state model, again a non-explicit function for $\gamma_{SV}$:

$$2\sqrt{\gamma_{SV}/\gamma_{LV}}\, e^{-\beta_1(\gamma_{LV}-\gamma_{SV})^2} - 1 - \cos\theta = 0 \qquad (7)$$

In equation 7, $\beta_1$ is a constant determined together with $\gamma_{SV}$ from a set of data of $\gamma_{LV}$ and $\theta$ measured on a solid surface, which is adjusted by a minimum square method. The average value obtained for $\beta_1$ is $0.0001247\ (m^2/mJ)^2$ [12].

By also using a theoretical approach, Kwok and Neumann [16] obtained a different equation of state relation as:

$$\gamma_{SL} = \gamma_{LV} + \gamma_{SV} - 2\sqrt{\gamma_{LV}\gamma_{SV}}\,(1 - B_2(\gamma_{LV} - \gamma_{SV})^2) \qquad (8)$$

which, combined with equation 1, results in the third Neumann equation of state relation, which is also a non-explicit function of $\gamma_{SV}$:

$$2\sqrt{\gamma_{SV}/\gamma_{LV}}\,(1 - \beta_2(\gamma_{LV} - \gamma_{SV})^2) - 1 - \cos\theta = 0 \qquad (9)$$

where $\beta_2$ is a constant determined using the same procedure used for $\beta_1$, resulting in an average value of $0.0001057\ (m^2/mJ)^2$ [16].

Equations 5, 7 and 9 are three different models of the Neumann equation of state which may be used to determine the surface free energy of a solid surface from contact angle measurements of only one probe liquid on the solid surface of interest. This is a great advantage of the Neumann method with respect to other methods which require the measurement of a set of data of $\theta$ for two or more different liquids [8,17–21]. Kwok and Neumann [12,16] have suggested that any of the three Neumann´s model given by equations 5, 6 and 9 could be used because they have obtained similar values of $\gamma_{SV}$ with the three models. This conclusion is based on numerical comparisons made using values of $\gamma_{LV}$ of 30 mJ/m², 50 mJ/m² and 70 mJ/m² and contact angles from 20° to 110° with increments of 10°.

It is worth mentioning that the Neumann equation of state method is not exempt from critics who question its usefulness to correctly estimate the values of surface free energy in solids [22–28].

In general, the error in the measurement of $\theta$ is significant [2,7,29], in the order of 1° to 4° [30–33] but its effect on the Neumann's model has not been determined or estimated. In addition to the experimental uncertainty of the determination of θ, the phenomenon of hysteresis in the contact angle of liquids deposited on real surfaces (i.e., rough and heterogeneous) must be considered [16,29]. Contact angle hysteresis describes the experimentally observed situation where a range of θ values is observed when depositing a drop on a solid substrate instead of a single equilibrium value as predicted by Young's equation [34–36]. The maximum observed value is called the advancing contact angle ($\theta_a$) and the minimum observed value is the receding contact angle ($\theta_r$), These contact angles can be obtained experimentally by different techniques (for example: tilting-plate method) [34]. Kwok and Neumann suggest that to calculate the surface free energy the advancing contact angle should be used since it is a very good approximation of the equilibrium contact angle and is highly reproducible [16] however there is no consensus on which angle value to use especially when the values of $\theta_r$ and $\theta_a$ are very different [37]. Therefore, a sensitivity analysis of the error in $\theta$ in the calculation of $\gamma_{SV}$ is necessary to establish the uncertainty of the value of $\gamma_{SV}$ calculated.

To calculate $\gamma_{SV}$ from the values of $\theta$ measured, the analytical functions given by equations 5, 7 and 9 require a procedure to calculate the roots of such functions appropriately adapted. Such roots must be calculated by numerical methods since there are no known analytical methods that could be applied [10,14]. A suitable numerical method is the Newton method [38]. To simplify the use of a numerical method and considering the complexity of the functions involved in equations 5, 7 and 9, it may be convenient to use an explicit function obtained from these equations. These can be obtained through a polynomial fit $P_{\gamma_{SV}}(\theta)$ to the data $\gamma_{SV}$ vs $\theta$ with the value of $\gamma_{SV}$ obtained by numerical methods from equation 9 (third and most recent equation of state model). Any uncertainty in the measurement of $\theta$ will also have impact on the error in $\gamma_{SV}$ and the polynomial obtained will allow the calculation and analysis of the propagation of uncertainties in an efficient way. In addition, it must be considered that the values of $\gamma_{LV}$ used in the calculations are generally taken from published tables. However, these may not strictly apply to the liquid used in the experiments, because the liquids used could have different degree of purity, some grade of contamination, or even some degree of degradation [28,29]. The value of $\gamma_{LV}$ is normally dependent on temperature [39,40], and, in some liquids, it may be relevant; for instance, the value of $\gamma_{LV}$ of water at 20 °C is 72.75±0.36 mJ/m$^2$, while that at 15 °C is 73.50±0.37 mJ/m$^2$ and at 25 °C is 71.99±0.36 mJ/m$^2$ [41,42], that is a change of ±5 °C leads to a variation in $\gamma_{LV}$ of ±0.75 mJ/m$^2$, which is near 1 %. Both uncertainties in $\theta$ and $\gamma_{LV}$ from different origins must be considered when determining a value of $\gamma_{SV}$. Therefore, it results necessary to determine how both uncertainties propagate on $\gamma_{SV}$. One way of doing this is finding a polynomial expansion $P_{\gamma_{SV}}(\theta, \gamma_{LV})$ from equations 5, 7 and 9 for the direct calculation of $\gamma_{SV}$ as a function of $\theta$ and $\gamma_{LV}$, i.e. an explicit function, and perform the partial derivates of $P_{\gamma_{SV}}$ with respect to $\theta$ and $\gamma_{LV}$ to calculate the uncertainty propagated in $\gamma_{SV}$ [43].

In the present study, the three Neumann models were used to calculate $\gamma_{SV}$ and the possible discrepancies among the models were analyzed. Also, a sensitivity analysis of each Neumann equation state model was performed based on the condition number [38]. Finally, polynomial fittings $P_{\gamma_{SV}}(\theta)$ and $PM_{\gamma_{SV}}(\theta, \gamma_{LV})$ were obtained to calculate

$\gamma_{SV}$ and were compared with the values calculated by numerical methods from equation 9.

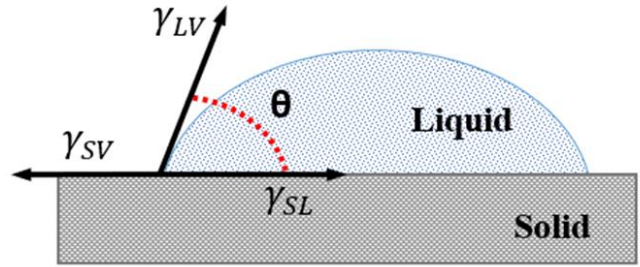

*Figure 1. Scheme of a sessile drop, contact angle ($\theta$), and the three interfacial tensions $\gamma_{LV}$, $\gamma_{SV}$ and $\gamma_{SL}$.*

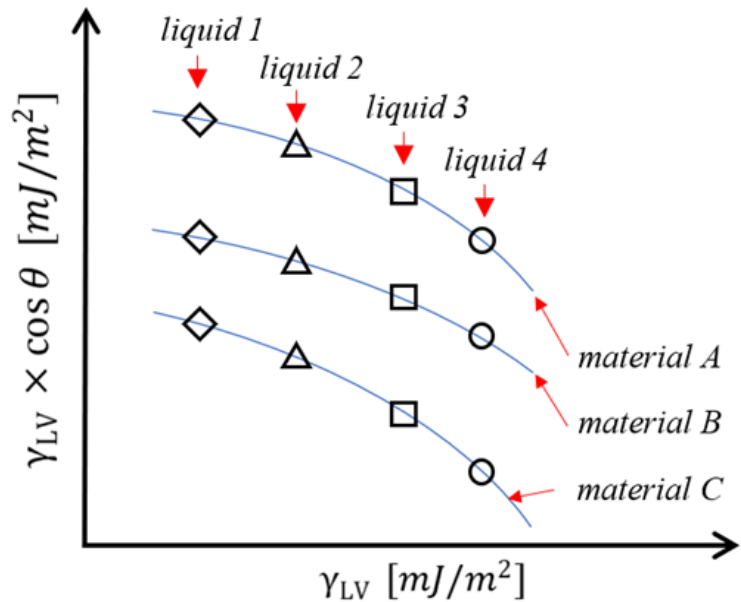

*Figure 2. Plot of $\gamma_{LV} \times \cos\theta$ versus $\gamma_{LV}$ for different liquids (1, 2, 3 and 4) on three different solid surfaces (A, B and C).*

## 2. Materials and Methods

First, the values of $\gamma_{SV}$ obtained by each of the equations of state relations were obtained and compared. Next, the sensitivity of each equation to changes in the value of contact angle was determined and, finally, two polynomial functions were obtained: one for the calculation of $\gamma_{SV}$ as a function of $\theta$ and another one for the calculation of $\gamma_{SV}$ as a function of $\theta$ and $\gamma_{LV}$.

## 2.1. Comparison of the values of $\gamma_{SV}$ obtained by the three models of the Neumann equation of state relation

Eleven values of $\gamma_{LV}$ of hypothetical liquids were used (25, 30, 35, 40, 45, 50, 55, 60, 65, 70 and 72.8 mJ/m²) and, for each of them, 130 values of $\theta$ from 1° to 130° with 1° increments were used to calculate $\gamma_{SV}$ from equations 5, 7 and 9. The $\gamma_{SV}$ values were obtained using the Newton numerical method [38,44], starting with an initial value of $\gamma_{SV}$ of 10 mJ/m² and stopping the iteration if $|g(\gamma_{SV})| < 10^{-10}$, where $g(\gamma_{SV})$ is equation 5, 7 or 9.

In the case of equation 5, due to the low values given by the denominator ($\gamma_{LV}(0.015\sqrt{\gamma_{LV}\gamma_{SV}} - 1)$), which tends to zero between 65 and 73 mJ/m² for $\gamma_{LV}$ and between 40 and 73 mJ/m² for $\gamma_{SV}$, a discontinuity is generated in the function. This discontinuity is analyzed in Appendix A. For comparison between the Neumann models, the $\gamma_{SV}$ for $\gamma_{LV}$ values of 65, 70 and 72.8 mJ/m² was numerically calculated for $\theta$ in the range of 111° to 130°. Then, the $\theta$ between 10° and 110° $\gamma_{SV}$ was obtained from Neumann et al. [45], and finally the $\theta$ between 1° and 9° $\gamma_{SV}$ was calculated using a polynomial interpolation of fifth order [14].

The values of $\gamma_{SV}$ obtained with equations 5, 7 and 9 were identified as $\gamma_{SV}^{I}$, $\gamma_{SV}^{II}$ and $\gamma_{SV}^{III}$, respectively. For each of the 11 values of $\gamma_{LV}$, 130 values of $\gamma_{SV}$ were calculated and therefore eleven $\gamma_{SV}(\theta)$ functions were obtained.

To compare among the values of $\gamma_{SV}$ obtained with each of the three models, the discrepancy $D^{i-k}$ and the relative percent discrepancy $RD^{i-k}$ were defined as:

$$D^{i-k} = \gamma_{SV}^{i} - \gamma_{SV}^{k} \tag{10}$$

$$RD^{i-k} = \left((\gamma_{SV}^{i} - \gamma_{SV}^{k})/\gamma_{SV}^{i}\right) \times 100\ \% = \left(D^{i-k}/\gamma_{SV}^{i}\right) \times 100\ \% \tag{11}$$

where $i$ and $k$ are the models compared. The units of $D^{i-k}$ are mJ/m² and $RD^{i-k}$ is dimensionless.

## 2.2. Sensitivity of the models to the contact angle

The sensitivity was analyzed through the calculation of the condition number $\mathbb{C}_{\gamma_{SV}}$ of the discrete functions $\gamma_{SV}(\theta)$ for each model of equation state relation calculated as [38]:

$$\mathbb{C}_{\gamma_{SV}} = \frac{\varepsilon_{\gamma_{SV}}}{\varepsilon_\theta} = \frac{(\gamma_{SV}(\theta+\Delta\theta)-\gamma_{SV}(\theta))/\gamma_{SV}(\theta)}{\Delta\theta/\theta} = \frac{\theta\left(\gamma_{SV}(\theta+\Delta\theta)-\gamma_{SV}(\theta)\right)}{\Delta\theta \times \gamma_{SV}(\theta)} \qquad (14)$$

where $\varepsilon_{\gamma_{SV}}$ and $\varepsilon_\theta$ are the relative errors in $\gamma_{SV}$ and $\theta$, respectively, and $\theta$ takes values from 1° to 130° with $\Delta\theta=1°$. If $\mathbb{C}_{\gamma_{SV}}<1$, the function is well conditioned, whereas if $\mathbb{C}_{\gamma_{SV}}>1$, the function is poorly or bad conditioned, indicating in this case that any $\varepsilon_\theta$ will magnify $\varepsilon_{\gamma_{SV}}$ [38]. It is noted that the condition number in equation 14 is also calculated as a function of $\theta$.

## 2.3. Polynomial fitting of $\gamma_{SV}$ as a function of $\theta$

A polynomial fitting of $\gamma_{SV}$ as a function of $\theta$ will allow calculating it directly without applying numerical iterative methods. Since the polynomial is only a function of $\theta$, the value of $\gamma_{LV}$ must be known; that is, for each $\gamma_{LV}$, there will be one specific polynomial.

In the present research, only the third and most recent model of equation of state relation was fitted by a polynomial, and the values of $\gamma_{SV}$ calculated with it were compared with those obtained by numerical calculation of the analytical function given by equation 9.

In the first polynomial fitting, the probe liquid used was water, the base data set of $\gamma_{SV}^{III}$ vs $\theta_w$ was obtained using equation 9, and, with this set, polynomials of $n^{th}$ order $P_n^w \gamma_{SV}(\theta)$ with n=2, 3, 4 and 5 were adjusted using the minimum square method. The value of $\gamma_{LV}$ used for water was 72.8 mJ/m² [46]. After this, the discrepancy between the polynomial results and the iterative numerical values of $\gamma_{SV}^{III}$ was determined and analyzed.

Once the analysis for water was completed, polynomial fittings of fifth order were obtained for the other probe liquids: glycerol, formamide, methylene iodide, ethylene glycol, 1-bromonaphthalene and dimethyl sulfoxide, using $\gamma_{LV}$ values of 65.02 mJ/m², 59.08 mJ/m², 49.98 mJ/m², 47.55 mJ/m², 44.31 mJ/m² and 42.68 mJ/m² respectively [47]. The discrepancy between the $\gamma_{SV}$ values given by the fitted polynomials and the $\gamma_{SV}^{III}$ values was analyzed as for the case of water.

## 2.4. Fifth order polynomial fitting of $\gamma_{SV}$ as a function of $\gamma_{LV}$ and $\theta$

The polynomial fitting was expanded to include not only the contact angle but also the surface tension between liquid and vapor of any probe liquid as $PM\gamma_{SV}(\theta, \gamma_{LV})$, approximating $\gamma_{SV}$ as a function of $\theta$ and $\gamma_{LV}$. In this case, as before, the polynomial of fifth order was obtained using the third model given by equation 9; that is, $\gamma_{SV}^{III}(\theta, \gamma_{LV}) \cong PM\gamma_{SV}(\theta, \gamma_{LV})$, where:

$$PM\gamma_{SV}(\theta, \gamma_{LV}) = A(\gamma_{LV}) \times \theta^5 + B(\gamma_{LV}) \times \theta^4 + C(\gamma_{LV}) \times \theta^3 + D(\gamma_{LV}) \times \theta^2 + E(\gamma_{LV}) \times \theta^1 + F(\gamma_{LV}) \tag{15}$$

where the coefficients A, B, C, D, E and F are also polynomials fitted as described in Appendix B.

In summary, first, 11 fifth-order polynomials were fitted for each value of $\gamma_{LV}$ corresponding to each hypothetical liquid probe, with respect to the data of $\gamma_{SV}^{III}$ vs $\theta$; and then, the coefficients A, B, C, D, E and F were fitted using the coefficients of the eleven polynomials for each $\gamma_{LV}$, that is, $P_5\gamma_{SV}(\theta)$ vs $\gamma_{LV}$.

## 3. Results and Discussion

### 3.1. Comparison of the values of $\gamma_{SV}$ obtained by the different Neumann equations of state

Figure 3 shows the graph of the 11 functions $\gamma_{SV}^{III}(\theta)$, one for each hypothetical probe liquid, obtained from the numerical resolution of equation 9. Each of the functions consists of 130 values of $\gamma_{SV}$, one for the value of each contact angle tested. It can be seen that the functions $\gamma_{SV}^{III}(\theta)$ monotonically decrease with $\theta$. The functions obtained from equations 5 ($\gamma_{SV}^{I}(\theta)$) and 7 ($\gamma_{SV}^{II}(\theta)$) show a similar behavior (graphs not shown); however, as shown below, the value of $\gamma_{SV}$ for a given pair of values of θ and $\gamma_{LV}$ differs from one equation of state or model to another.

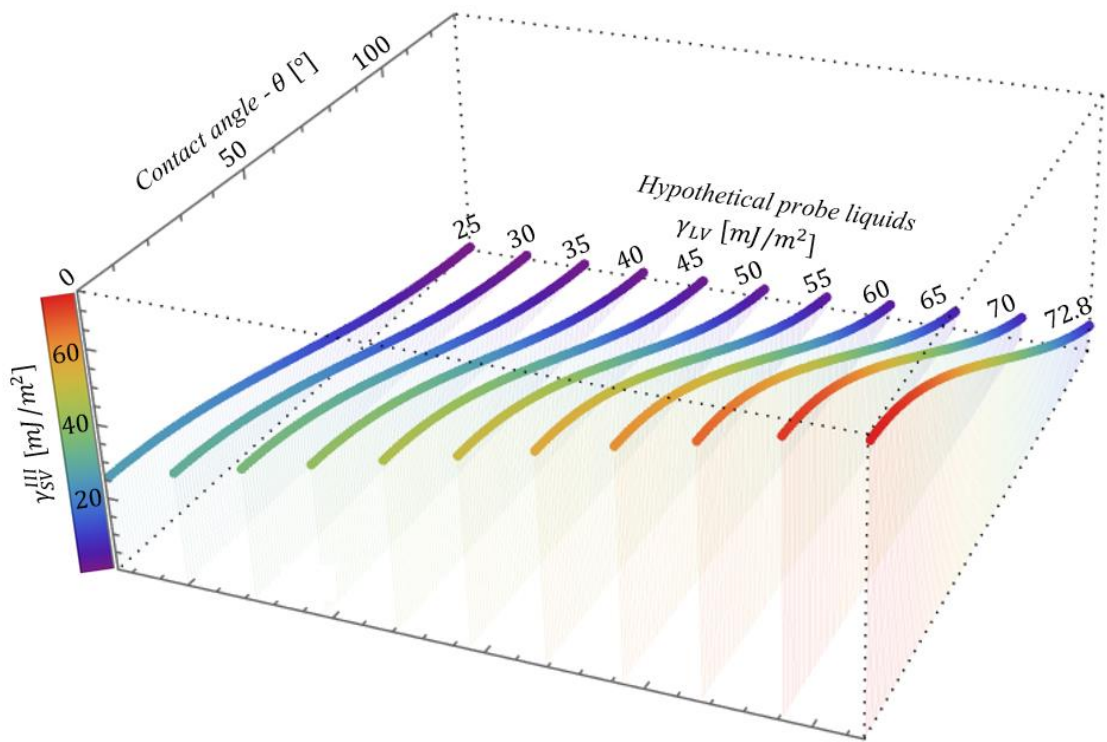

*Figure.3. Values of $\gamma_{SV}^{III}$ as a function of θ, obtained for each of the 11 hypothetical probe liquids ($\gamma_{LV}$). The fill marks the distance between $\gamma_{SV}^{III} = 0\ mJ/m^2$ and the point corresponding to each value of $\gamma_{SV}^{III}$ calculated.*

First, a comparative global analysis of the 1430 values (11 liquid × 130 θ values) of $\gamma_{SV}$ obtained for each combination of θ and $\gamma_{LV}$ with each of the equations of state was performed analyzing the values of discrepancy ($D$) and relative discrepancy ($RD$). The results of the accumulated empirical distribution functions of the discrepancy values $D^{III-I}$, $D^{III-II}$ and $D^{II-I}$ are shown in Figure 4.A, while the empirical cumulative distribution functions of the relative discrepancy values $RD^{III-I}$, $RD^{III-II}$ and $RD^{II-I}$ are shown in Figure 4.B.

The maximum values of $D^{III-I}$, $D^{III-II}$ and $D^{II-I}$ found were 1.28 $mJ/m^2$, 0.78 $mJ/m^2$ and 0.90 $mJ/m^2$, while the minimum values were -0.65 $mJ/m^2$, -0.66 $mJ/m^2$ and -0.25 $mJ/m^2$, respectively. It can be seen that 90% of the central values (determined by the 95th percentile and the 5th percentile of the data) are between 0.47 and -0.30 $mJ/m^2$ for $D^{III-I}$, 0.05 and -0.54 $mJ/m^2$ for $D^{III-II}$, and 0.73 and -0.20 $mJ/m^2$ for $D^{II-I}$. The low magnitude of both the maximum and minimum values of the discrepancy and the values of the 90% interval of the central data indicate that the three equations provide generally similar values of $\gamma_{SV}$; however, this analysis does not consider the magnitude of the values of $\gamma_{SV}$. To take this magnitude into account, it is necessary to analyze the relative discrepancy. In such case, the maximum values of $RD^{III-I}$, $RD^{III-II}$ and $RD^{II-I}$ were 15.8%, 10.1% and 6.4%, while the minimum values were -16.4%, -2.8% and -13.9%, respectively. In this case, it can be seen that 90% of the central values are between 3.4 and -8.6 $mJ/m^2$ for $RD^{III-I}$, 0.4 and -2.4 $mJ/m^2$ for $RD^{III-II}$, and 3.8 and -6.2 $mJ/m^2$ for $RD^{II-I}$. These results show that the RD values are not negligible. Next, we analyzed in detail with which values of $\gamma_{SV}$ the values of $RD$ are associated.

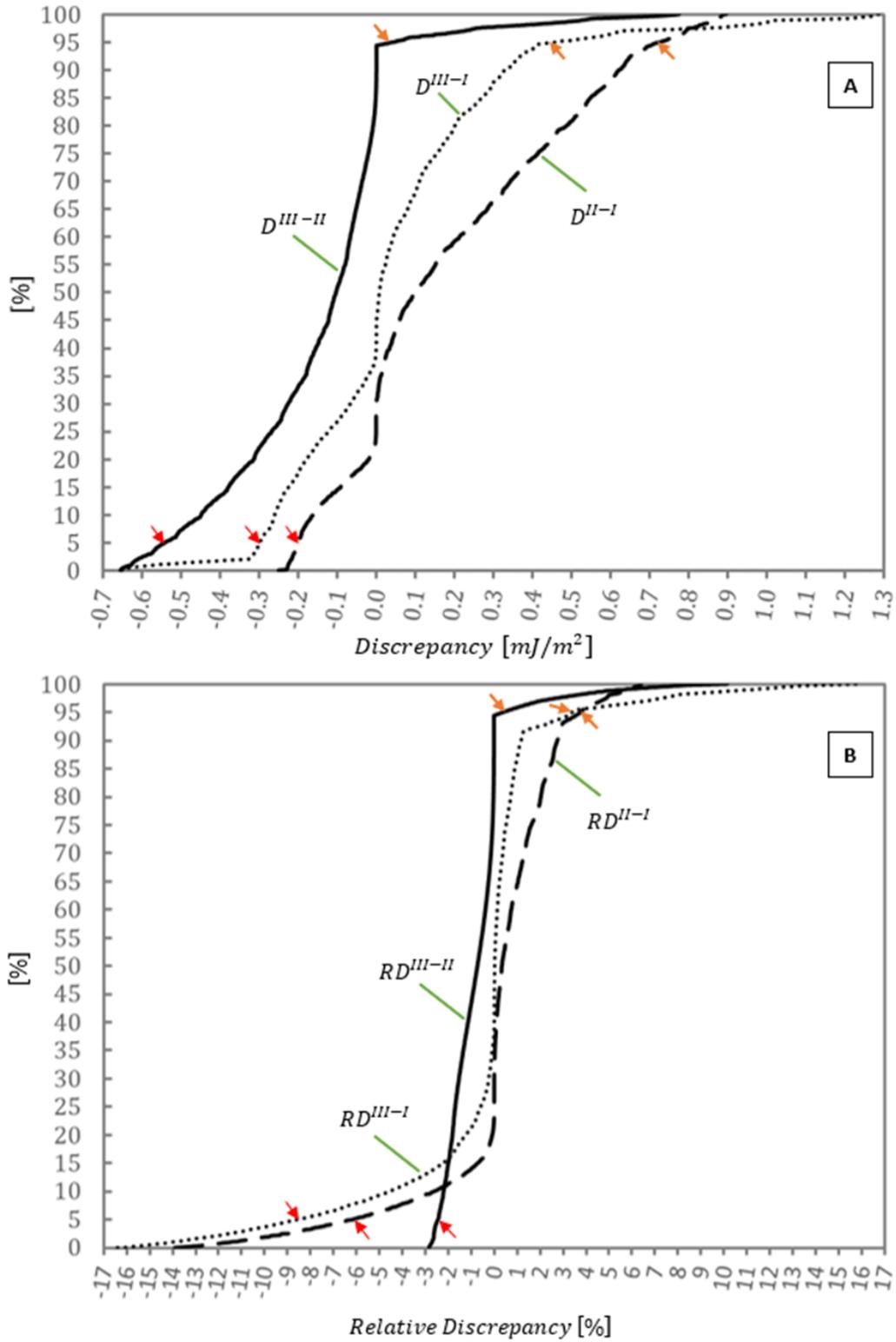

*Figure 4. Cumulative empirical distribution functions of discrepancy (A) and relative discrepancy (B). The red arrows indicate the 5th percentile, while the orange arrows indicate the 95th percentile.*

To analyze the relationship between the values of $RD$ and the values of $\gamma_{SV}$, the behavior of $RD$ as a function of $\theta$ and $\gamma_{LV}$ was studied using level curves. Figures 5, 6 and 7 show the level curves for $RD^{III-I}, RD^{III-II}$ and $RD^{II-I}$, respectively. The three figures show that both the maximum and minimum values of $RD$ are obtained when the value of $\theta$ is high (~100°), that is, when the value of $\gamma_{SV}$ is lower. In general, the $\gamma_{SV}$ values reported in the literature for different materials are greater than ~10 $mJ/m^2$ [48–52], so it is of practical interest to analyze $RD^{III-I}, RD^{III-II}$ and $RD^{II-I}$ for the case where $\gamma_{SV}$ is equal to or greater than that value. As a reference for the analysis, in Figures 5, 6 and 7, the combinations $\theta$ and $\gamma_{LV}$ are marked by means of a continuous green line when they result in a value of $\gamma_{SV}$ =10 $mJ/m^2$, by means of a light blue line when they result in a value $\gamma_{SV}$ =25 $mJ/m^2$ and by means of a red line when they result in a value of $\gamma_{SV}$ =50 $mJ/m^2$. These graphs represent a map that allows analyzing the influence of the probe liquid ($\gamma_{LV}$ = constant) and the unknown solid material ($\gamma_{SV}$ = constant) in the $RD$ and detecting areas of greater or lesser differences in the models. In subsequent analyses, materials with $\gamma_{SV} \geq 10$ $mJ/m^2$ will be considered.

Figure 5 shows that when $\gamma_{SV} \cong 10$ $mJ/m^2$, the $RD^{III-I}$ values vary considerably (between -3% and 10%) depending on the probe liquid used, whereas when $\gamma_{SV} \cong 25$ $mJ/m^2$, the values of $RD^{III-I}$ vary between -0.1% and +1%. On the other hand, when $\gamma_{LV}$ takes intermediate values (45 $mJ/m^2$ to 55 $mJ/m^2$), the values of $RD^{III-I}$ vary from -1% to 1%, when $\gamma_{LV}$ is between 25 $mJ/m^2$ and 45 $mJ/m^2$, the values of $RD^{III-I}$ vary from -1% to -2%, and when the values of $\gamma_{LV}$ are greater than 55 $mJ/m^2$, the values of $RD^{III-I}$ vary from 1% to 10%.

Figure 6 shows that when $\gamma_{SV} \cong 10$ $mJ/m^2$, the values of $RD^{III-II}$ also vary appreciably between -1% and 8% depending on the probe liquid used. In contrast, when $\gamma_{SV} \cong 25$

$mJ/m^2$, the values of $RD^{III-I}$ vary between -0.1% and -1%. On the other hand, when $\gamma_{LV}$ takes intermediate values (60 mJ/m2 to 65 $mJ/m^2$), the values of $RD^{III-II}$ vary from -1% to 1%, when $\gamma_{LV}$ is between 25 $mJ/m^2$ and 60 $mJ/m^2$, the values of $RD^{III-II}$ vary from -0.1% to -2%, and when the values of $\gamma_{LV}$ are greater than 65 $mJ/m^2$, the $RD^{III-II}$ values vary from 1% to 8%.

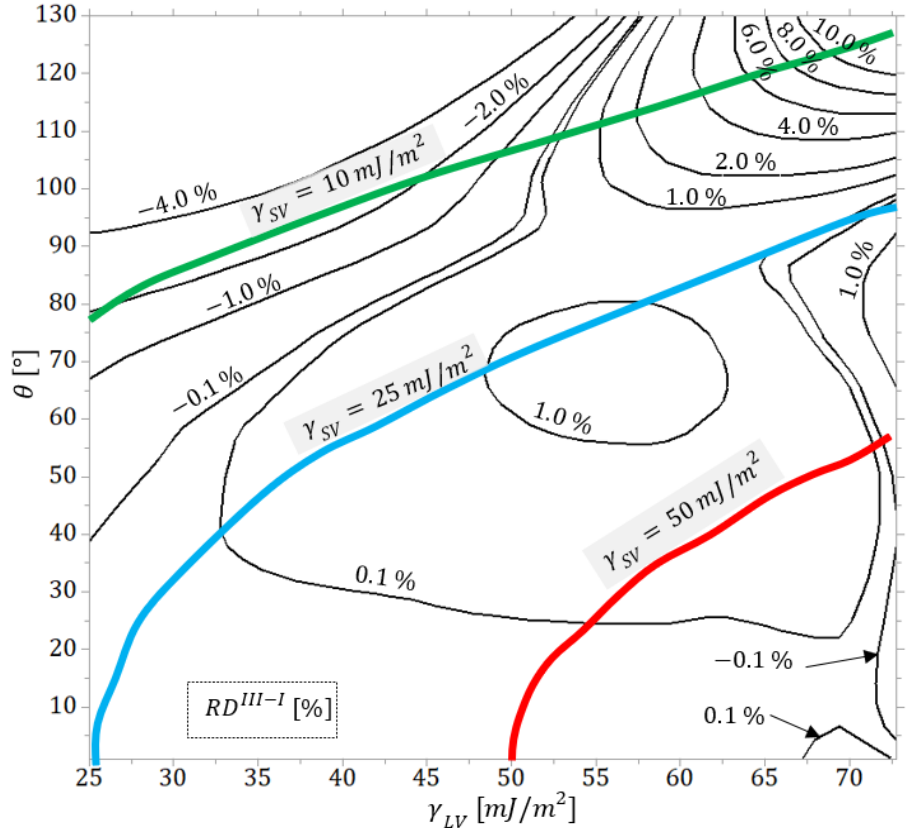

Figure 5. Level curve graph of $RD^{III-I}$ as a function of $\theta$ and $\gamma_{LV}$. The colored lines indicate a given value of $\gamma_{SV}$ (green: $\gamma_{SV}$=10 mJ/m², light blue: $\gamma_{SV}$=25J/m², red: $\gamma_{SV}$=50 mJ/m²). The graph is constructed by Akima interpolation [53] from 11×130=1430 values of $RD^{III-I}$.

Finally, Figure 7 shows that when $\gamma_{SV} \cong 10$ mJ/m², the $RD^{III-I}$ values vary appreciably depending on the probe liquid used. On the other hand, when $\gamma_{LV}$ takes values between 25 mJ/m² and 45 $mJ/m^2$, the $RD^{III-I}$ values vary from -1% to 1%, while when $\gamma_{LV}$ is greater than 45 $mJ/m^2$, the values of $RD^{III-I}$ vary from 1% to 6%.

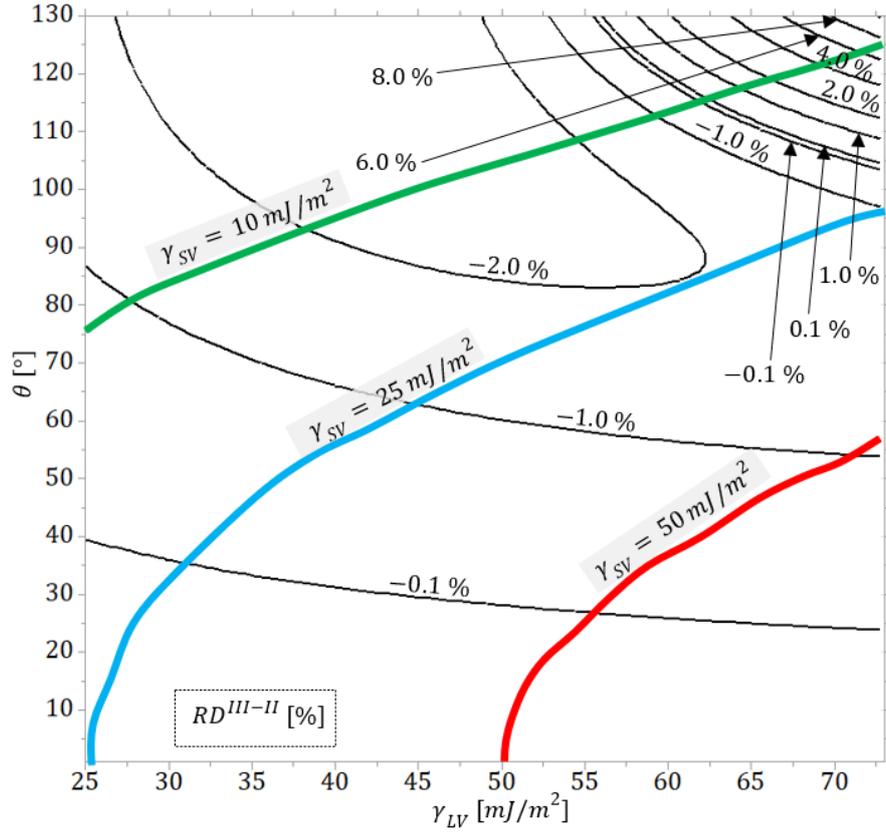

*Figure 6. Level curve graph of $RD^{III-II}$ as a function of $\theta$ and $\gamma_{LV}$. The colored lines indicate a given value of $\gamma_{SV}$ (green: $\gamma_{SV}$=10 mJ/m², light blue: $\gamma_{SV}$=25J/m², red: $\gamma_{SV}$=50 mJ/m²). The plot is constructed by Akima interpolation [53] from 11×130=1430 values of $RD^{III-II}$.*

In all three cases, it can be seen that, as the value of $\gamma_{SV}$ increases, the value of $RD$ decreases appreciably, for example, when $\gamma_{SV} \cong 50 \ mJ/m^2$, the values of $RD^{III-I}$, $RD^{III-II}$ and $RD^{II-I}$ vary from -2% to 2% regardless of the value of $\gamma_{LV}$ of the probe liquid used.

### 3.2. Sensitivity of the equations of state to the uncertainty in the contact angle

To analyze the sensitivity of the value of $\gamma_{SV}$ calculated by means of the equations of state to the uncertainty in the determination of $\theta$, we analyzed the condition number $(\mathbb{C}_{\gamma_{SV}}(\theta))$ expressed as a function of the contact angle. For each equation of state, 11 discrete functions (one for each value of $\gamma_{LV}$) of the condition number of $\gamma_{SV}(\theta)$ versus the contact angle ($\mathbb{C}_{\gamma_{SV}}(\theta)$) were obtained. For practical purposes, it is more useful to analyze the function $\mathbb{C}_{\gamma_{SV}}(\gamma_{SV})$, that is, the condition number of the equation of state with respect to $\theta$ when it takes a given value of $\gamma_{SV}$ to analyze the effect of different test liquids used on the same solid material. The transition from one function to another is trivial since, for a given value of $\gamma_{LV}$, each value of $\theta$ corresponds to a value of $\gamma_{SV}$ determined by the Neumann equation of state used.

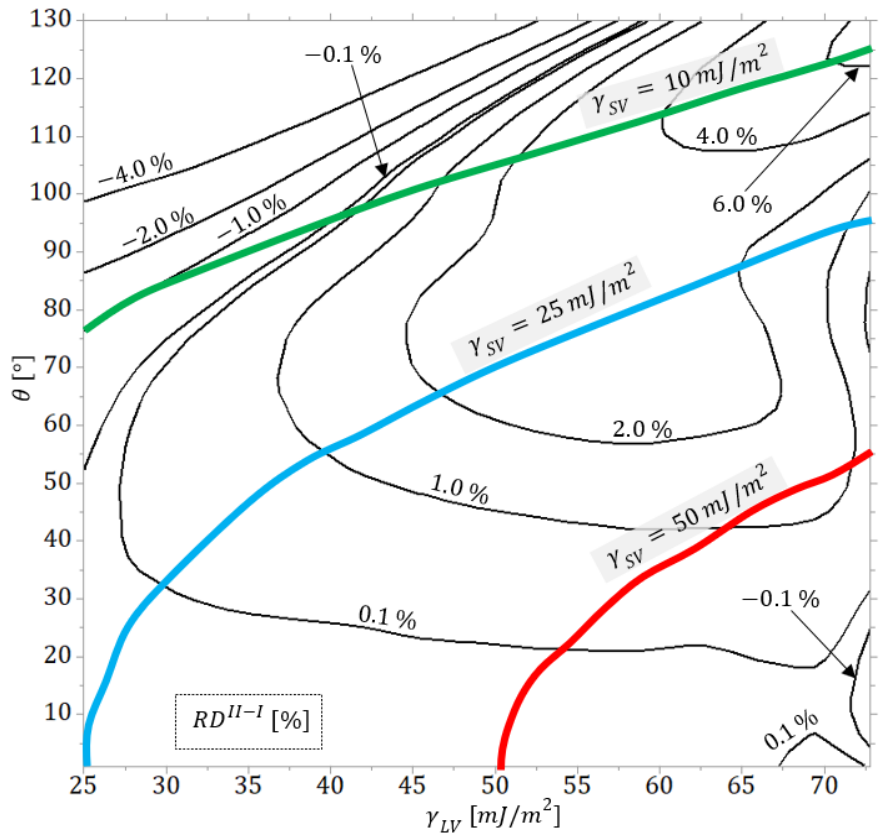

Figure 7. Level curve graph of $RD^{II-I}$ as a function of $\theta$ and $\gamma_{LV}$. The colored lines indicate a given value of $\gamma_{SV}$ (green: $\gamma_{SV}=10$ mJ/m², light blue: $\gamma_{SV}=25$ mJ/m², red: $\gamma_{SV}=50$ mJ/m²). The plot is constructed by Akima interpolation [53] from 11×130=1430 values of $RD^{II-I}$.

Figure 8.A shows the functions $\mathbb{C}_{\gamma_{SV}^{III}}(\gamma_{SV}^{III})$, where, for clarity, only the cases in which $\gamma_{LV}$ takes the values of 25, 35, 45, 55, 65 and 72.8 $mJ/m^2$ are shown, since the functions for all the other cases follow the same pattern. It can be seen that, for a given

value of $\gamma_{SV}$, the condition number is lower when a liquid with lower $\gamma_{LV}$ is used. In general, it is true that $\mathbb{C}_{\gamma_{SV}^{III}}(\gamma_{SV}^{III})<1$ in the range from $\gamma_{SV}^{max}$ (maximum value of $\gamma_{SV}$ that can be calculated with a particular probe liquid $\gamma_{SV}^{max} = \gamma_{LV}$ when $\theta= 0°$) to about 60% of $\gamma_{SV}^{max}$. In the range from 60% of $\gamma_{SV}^{max}$ to $\gamma_{SV}^{min}$ ($\gamma_{SV}^{min}$ being the value of $\gamma_{SV}$ when $\theta=130°$), the value of $\mathbb{C}_{\gamma_{SV}^{III}}(\gamma_{SV}^{III})$ increases until reaching values ranging from ~7 at $\gamma_{LV} = 72.8 \ mJ/m^2$ to ~9 at $\gamma_{LV} = 25 \ mJ/m^2$. The behavior of the functions $\mathbb{C}_{\gamma_{SV}^{I}}(\gamma_{SV}^{I})$ and $\mathbb{C}_{\gamma_{SV}^{II}}(\gamma_{SV}^{II})$ (not shown in the figure) is very similar to that of $\mathbb{C}_{\gamma_{SV}^{III}}(\gamma_{SV}^{III})$.

The discrete function $\mathbb{C}_{\gamma_{SV}}(\theta)$ allowed obtaining a relation which in turn allowed obtaining the behavior of $\varepsilon_{\gamma_{SV}}$ with respect to $\theta$ and the absolute error $E_\theta$ y as:

$$\varepsilon_{\gamma_{SV}} = \mathbb{C}_{\gamma_{SV}}(\theta) \times \varepsilon_\theta = \mathbb{C}_{\gamma_{SV}}(\theta) \times E_\theta/\theta \qquad (16)$$

Note that $\varepsilon_{\gamma_{SV}}$ will be a function of both $\theta$ and $E_\theta$, that is $\varepsilon_{\gamma_{SV}}(\theta, E_\theta)$. The absolute error $E_\theta$ can be considered as the standard uncertainty in the experimental determination of the contact angle [54,55] and, for this analysis, the particular case in which the standard uncertainty is ±1° was considered, for which $E_\theta = 1°$. As in the previous case, it is more useful to analyze the behavior of the relative error in $\gamma_{SV}$ when it is a function of the particular values taken by $\gamma_{SV}$, that is, the function $\varepsilon_{\gamma_{SV}}(\gamma_{SV})$.

Figure 8.B shows the functions $\varepsilon_{\gamma_{SV}^{III}}(\gamma_{SV}^{III})$ expressed in percentages, that is, the percentage relative error in $\gamma_{SV}^{III}$ when $\gamma_{SV}^{III}$ takes a given value considering that $E_\theta=1°$, where, for clarity, only the cases in which $\gamma_{LV}$ takes the values of 25, 35, 45, 55, 65 and 72.8 $mJ/m^2$ are shown, since the functions for all the other cases follow the same pattern. The analysis of Figure 8.B allows deducing that if there is an experimental system to measure $\theta$ whose minimum standard uncertainty is ±1°, the lower bounds of relative error in $\gamma_{SV}^{III}$ are those shown in that figure. As can be seen, these bounds are different for each probe liquid. For example, if there is a solid surface with a value of $\gamma_{SV} \cong 20\ mJ/m^2$, when determining the $\gamma_{SV}$ by means of the contact angle of a probe liquid whose $\gamma_{LV}$ is $25\ mJ/m^2$, the result will have a minimum uncertainty of ~1%, whereas, if we do it with water ($\gamma_{LV} = 72.8\ mJ/m^2$), the result will have a minimum uncertainty of ~3%. It is possible to analyze other situations where $E_\theta \neq 1°$ simply by multiplying the values shown in Figure 8.B by the value of $E_\theta$ hypothetical standard uncertainty to be considered (in degrees).

### 3.3. Polynomial fit of $\gamma_{SV}$ as a function of $\theta$

From the values of $\gamma_{SV}^{III}$ obtained by numerical resolution of equation 9 and considering water as the probe liquid, fit polynomials (of degree 2, 3, 4 and 5) were obtained based on the contact angle values of water ($\theta_w$). The polynomials found are shown below:

$$P_2^w \gamma_{SV}(\theta_w) = -7.8318 \times 10^{-4} \theta_w^2 - 4.5257 \times 10^{-1} \theta_w + 7.7092 \times 10^1 \qquad (17)$$

$$P_3^w \gamma_{SV}(\theta_w) = 2.8628 \times 10^{-5} \theta_w^3 - 6.4086 \times 10^{-3} \theta_w^2 - 1.5667 \times 10^{-1} \theta_w + 7.3800 \times 10^1 \qquad (18)$$

$$P_4^w \gamma_{SV}(\theta_w) = -1.6247 \times 10^{-7} \theta_w^4 + 7.1195 \times 10^{-5} \theta_w^3 - 1.0003 \times 10^{-2} \theta_w^2 - 5.1101 \times 10^{-2} \theta_w + 7.3085 \times 10^1 \qquad (19)$$

$$P_5^w \gamma_{SV}(\theta_w) = 1.8826 \times 10^{-9} \theta_w^5 - 7.7902 \times 10^{-7} \theta_w^4 + 1.4313 \times 10^{-4} \theta_w^3 - 1.3557 \times 10^{-2} \theta_w^2 + 1.6449 \times 10^{-2} \theta_w + 7.2774 \times 10^1 \qquad (20)$$

where the values of $\theta_w$ are in degrees (deg) and the units of the coefficients multiplying $\theta_w^5$, $\theta_w^4$, $\theta_w^3$, $\theta_w^2$, and $\theta$ are $mJ/(m^2\ deg^5)$, $mJ/(m^2\ deg^4)$, $mJ/(m^2\ deg^3)$, $mJ/(m^2\ deg^2)$ and $mJ/(m^2\ deg)$, respectively. The units of the independent term are mJ/m². The domain of the polynomial functions is *1° ≤ θ_W ≤ 130°* and the functions are valid only when water is used as the probe liquid. The coefficients are expressed with five significant digits to reach the tolerance expressed by the relative discrepancy reported in Table 1 [38].

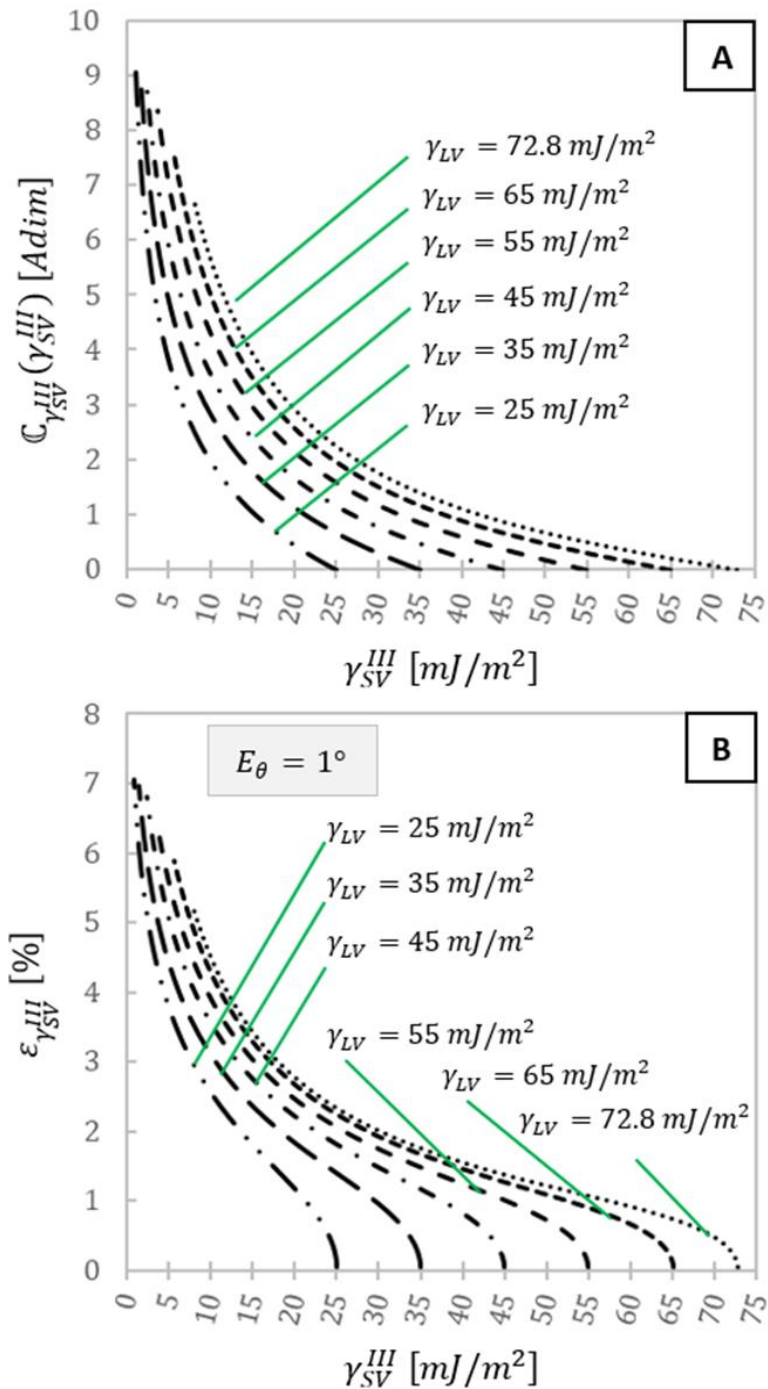

Figure 8. A-) Condition number of the Neumann equation of state ($\mathbb{C}_{\gamma_{SV}^{III}}$) as a function of $\gamma_{SV}^{III}$. B-) Percentage relative error in $\gamma_{SV}^{III}$ ($\varepsilon_{\gamma_{SV}^{III}}$) as a function of $\gamma_{SV}^{III}$ considering an absolute error in $\theta$ of 1°.

Figure 9 shows the relative discrepancy between the values of $\gamma_{SV}^{III}$ and the values of $\gamma_{SV}$ obtained with the polynomials of degree 2, 3, 4 and 5 *vs.* $\gamma_{SV}^{III}$. It can be seen that the relative discrepancy decreases appreciably as the degree of the polynomial increases, until it remains at values very close to ~0% in the case where the polynomial is of degree 5. Table 1 shows the root mean square, maximum value, minimum value, and the range, both of the discrepancy ($D$) and the relative discrepancy ($RD$) between the values of $\gamma_{SV}^{III}$ and the values of $\gamma_{SV}$ obtained with each of the polynomials. For the case of the discrepancy, it can be observed that the value of the root mean square $\bar{x}_{rms}^{D}$ decreases abruptly with the increase in the order of the fit polynomial. Thus, we can see that, taking the value of the $\bar{x}_{rms}^{D}$ of the 2nd order polynomial as a basis for comparison, the $\bar{x}_{rms}^{D}$ of the 3rd order polynomial is ~5 times smaller, the $\bar{x}_{rms}^{D}$ of the 4th order polynomial is ~14 times smaller, and the $\bar{x}_{rms}^{D}$ of the 5th order polynomial is ~170 times smaller. Something similar happens with the root mean square of the relative discrepancy $\bar{x}_{rms}^{D}$, i.e., in the comparison between the value of the $\bar{x}_{rms}^{D}$ of the 2nd order polynomial, the $\bar{x}_{rms}^{D}$ of the 3rd order polynomial is ~ 7 times smaller, the $\bar{x}_{rms}^{D}$ of the 4th order polynomial is ~14 times smaller and the $\bar{x}_{rms}^{D}$ of the 5th order polynomial is ~196 times smaller. In other words, the effort to use a higher order polynomial function is highly rewarded by the gain in accuracy of the result.

These polynomials can be compared with the one obtained by Deshmukh & Shetty [56] who presented a polynomial function of 3rd order obtained from the fit of the data of the numerical resolution of the Neumann equation of state (equation 7) vs. the water contact angle (from 0º to 150º), in which the coefficients of $\theta^3$, $\theta^2$, and $\theta$ reported are 2.9×10$^{-5}$, -0.00652 and -0.1326 and the independent term is 72.8. These values are similar to those of the polynomial function $P_3^w \gamma_{SV}(\theta)$ found in this work. The differences observed are mainly due to the fact that: i) Deshmukh & Shetty used the second Neumann model and in the present research the most modern state equation is adjusted (the third Neumann model, equation 9), ii) the range of contact angles used in the work by Deshmukh & Shetty and the present study is different, and iii) the coefficients of the polynomial function have less significant digits than in the present research. Deshmukh & Shetty did not report the magnitude of the discrepancies of the polynomial function compared to the values obtained by numerical resolution. However, analyzing the polynomial function $P_3^w \gamma_{SV}(\theta)$ obtained in the present work, it can be observed that the relative discrepancy values of this function are non-negligible, between -4.995% and 1.131%, while the discrepancy values are between -0.848 mJ/m² and 0.385 $mJ/m^2$.

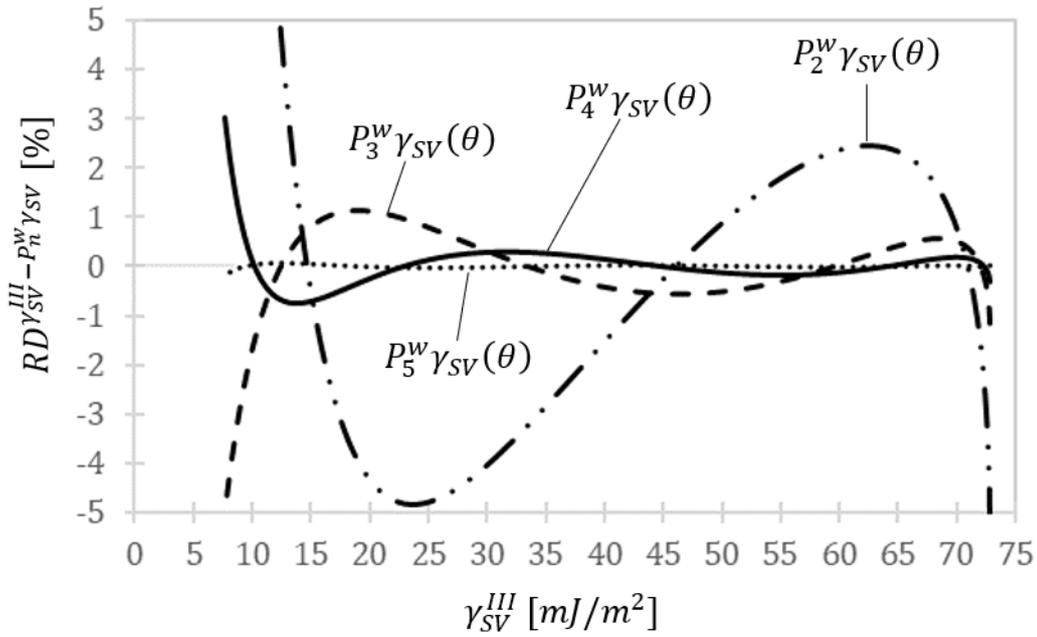

Figure 9. Behavior of the relative discrepancy (RD) between the $\gamma_{SV}^{III}$ values and the $\gamma_{SV}$ values obtained through the fit polynomials of degree 2, 3, 4 and 5. Ref.: $RD^{\gamma_{SV}^{III} - P_n^w \gamma_{SV}} = ((\gamma_{SV}^{III} - P_n^w \gamma_{SV})/\gamma_{SV}^{III}) \times 100\%$.

On the other hand, Żenkiewicz [17] presented three 3rd order polynomials obtained by fitting the Neumann equation of state (equation 7) solved by numerical methods for three probe liquids (water, formamide and methylene iodide) for contact angles between 0° and 125°. However, the values of surface free energy calculated from these polynomials are substantially different from those obtained in this work by numerical resolution of equation 7 as well as from those calculated by the polynomials proposed in this work and those proposed by Deshmukh and Shetty [56].

Table 1. Discrepancy (D) and relative discrepancy (RD) between $\gamma_{SV}^{III}$ and $\gamma_{SV}$ values obtained by fitting polynomials using water as probe liquid.

| Discrepancy $D^{\gamma_{SV}^{III} - P_n^w \gamma_{SV}} = \gamma_{SV}^{III} - P_n^w \gamma_{SV}$ [mJ/m$^2$] | | | | |
|---|---|---|---|---|
| | $n = 2$ | $n = 3$ | $n = 4$ | $n = 5$ |
| $\bar{x}_{RMS}$ | 1.211 | 0.236 | 0.084 | 0.007 |
| $Max$ | 2.620 | 0.385 | 0.232 | 0.012 |
| $Min$ | -3.850 | -0.848 | -0.235 | -0.015 |
| $Range$ | 6.470 | 1.233 | 0.467 | 0.026 |
| Relative Discrepancy $RD^{\gamma_{SV}^{III} - P_n^w \gamma_{SV}} = ((\gamma_{SV}^{III} - P_n^w \gamma_{SV})/\gamma_{SV}^{III}) \times 100$ [%] | | | | |
| $\bar{x}_{RMS}$ | 6.255 | 0.939 | 0.458 | 0.032 |
| $Max$ | 34.281 | 1.131 | 3.037 | 0.070 |
| $Min$ | -5.289 | -4.995 | -0.762 | -0.193 |
| $Range$ | 39.570 | 6.126 | 3.800 | 0.263 |

$n$: *order of the polynomial function*, $\bar{x}_{RMS}$: *Root-Mean-Square*; $Max$: *maximum value*; $Min$: *minimum value*; $Range$: $Max - Min$

Considering that, when using water as probe liquid, the 5$^{th}$ order polynomial fit ($P_5^w \gamma_{SV}(\theta)$) had discrepancy and relative discrepancy values that were acceptably low, 5$^{th}$ order polynomial fits were carried out for the other probe liquids mentioned before. $\gamma_{SV}^{III}$ vs $\theta_i$ values were adjusted, where $i$ represents the probe liquids used here, which were: glycerol (G), formamide (F), methylene iodide (M), ethylene glycol (E), 1-bromonaphthalene (B) and dimethyl sulfoxide (D). In each case, the following equations were obtained:

$$P_5^G \gamma_{SV}(\theta_G) = +1.6313 \times 10^{-9} \theta_G{}^5 - 6.6215 \times 10^{-7} \theta_G{}^4 + 1.2574 \times 10^{-4} \theta_G{}^3 -$$
$$1.2429 \times 10^{-2} \theta_G{}^2 + 2.0890 \times 10^{-2} \theta_G + 6.4974 \times 10^1 \quad (21)$$

$$P_5^F \gamma_{SV}(\theta_F) = 1.3318 \times 10^{-9} \theta_F{}^5 - 5.4162 \times 10^{-7} \theta_F{}^4 + 1.0907 \times 10^{-4} \theta_F{}^3 -$$
$$1.1367 \times 10^{-2} \theta_F{}^2 + 2.1219 \times 10^{-2} \theta_F + 5.9030 \times 10^1 \quad (22)$$

$$P_5^M \gamma_{SV}(\theta_M) = 7.1349 \times 10^{-10} \theta_M^5 - 3.1685 \times 10^{-7} \theta_M^4 + 7.9452 \times 10^{-5} \theta_M^3 -$$
$$9.4685 \times 10^{-3} \theta_M^2 + 1.7539 \times 10^{-2} \theta_M + 4.9937 \times 10^1 \qquad (23)$$

$$P_5^E \gamma_{SV}(\theta_E) = 5.3477 \times 10^{-10} \theta_E^5 - 2.5481 \times 10^{-7} \theta_E^4 + 7.1351 \times 10^{-5} \theta_E^3 -$$
$$8.9324 \times 10^{-3} \theta_E^2 + 1.6083 \times 10^{-2} \theta_E + 4.7511 \times 10^1 \qquad (24)$$

$$P_5^B \gamma_{SV}(\theta_B) = 3.0142 \times 10^{-10} \theta_B^5 - 1.7495 \times 10^{-7} \theta_B^4 + 6.0839 \times 10^{-5} \theta_B^3 -$$
$$8.2148 \times 10^{-3} \theta_B^2 + 1.4065 \times 10^{-2} \theta_B + 4.4276 \times 10^1 \qquad (25)$$

$$P_5^D \gamma_{SV}(\theta_D) = 1.8909 \times 10^{-10} \theta_D^5 - 1.3684 \times 10^{-7} \theta_D^4 + 5.5755 \times 10^{-5} \theta_D^3 -$$
$$7.8555 \times 10^{-3} \theta_D^2 + 1.3057 \times 10^{-2} \theta_D + 4.2648 \times 10^1 \qquad (26)$$

where the values of $\theta_i$ are in degrees (deg) and correspond to the contact angle of the liquid in question. The units of the coefficients that multiply $\theta_i^5$, $\theta_i^4$, $\theta_i^3$, $\theta_i^2$ and $\theta_i$ are $mJ/(m^2\ deg^5)$, $mJ/(m^2 deg^4)$, $mJ/(m^2 deg^3)$, $mJ/(m^2 deg^2)$ and $mJ/(m^2 deg)$, respectively. The units of the independent term are $mJ/m^2$. The domain of the polynomial functions is 1°≤θ≤130° and the functions are valid only when the probe liquid for which they were adjusted is used.

Table 2 shows the root mean square, maximum value, minimum value, and the range, both of the discrepancy and of the relative discrepancy between the values of $\gamma_{SV}^{III}$ and the values of $\gamma_{SV}$ obtained with the adjusted polynomials for each of the probe liquids considered. In all cases, the values of $\bar{x}_{rms}^D$ are less than or equal to 0.007 $mJ/m^2$, and the discrepancy values are within the range from 0.031 $mJ/m^2$ to -0.020 $mJ/m^2$. While the values of $\bar{x}_{rms}^D$ are always less than 0.125%, the relative discrepancy values range from 0.291% to -0.983%.

The seven polynomials $P_5^i \gamma_{SV}(\theta)$ found for the seven probe liquids considered allow, in addition to the direct calculation of $\gamma_{SV}$, the propagation of the experimental uncertainty of $\theta$ quickly and easily by means of the Taylor series method [57,58].

Table 2. Discrepancy and relative discrepancy between the $\gamma_{SV}^{III}$ and $\gamma_{SV}$ values obtained by fifth-order fit polynomials for different probe liquids.

| Discrepancy $D^{\gamma_{SV}^{III} - P_5\gamma_{SV}} = \gamma_{SV}^{III} - P_5\gamma_{SV}\ [mJ/m^2]$ | | | | | | |
|---|---|---|---|---|---|---|
| | G | F | M | E | B | D |
| $\bar{x}_{RMS}$ | 0.005 | 0.006 | 0.007 | 0.007 | 0.007 | 0.006 |
| $Max$ | 0.027 | 0.031 | 0.027 | 0.025 | 0.022 | 0.020 |
| $Min$ | -0.014 | -0.015 | -0.019 | -0.020 | -0.020 | -0.020 |
| $Range$ | 0.041 | 0.047 | 0.046 | 0.045 | 0.042 | 0.040 |
| Relative Discrepancy $RD^{\gamma_{SV}^{III} - P_5\gamma_{SV}} = ((\gamma_{SV}^{III} - P_5\gamma_{SV})/\gamma_{SV}^{III}) \times 100\ [\%]$ | | | | | | |
| $\bar{x}_{RMS}$ | 0.022 | 0.036 | 0.090 | 0.105 | 0.120 | 0.125 |
| $Max$ | 0.042 | 0.088 | 0.224 | 0.259 | 0.290 | 0.291 |
| $Min$ | -0.163 | -0.268 | -0.703 | -0.817 | -0.935 | -0.983 |
| $Range$ | 0.205 | 0.356 | 0.927 | 1.075 | 1.224 | 1.273 |

$\bar{x}_{RMS}$: $Root\text{-}Mean\text{-}Square$; $Max$: $maximum\ value$; $Min$: $minimum\ value$; $Range$: $Max - Min$.
G: glycerol, F: formamide, M: methylene iodide, E: ethylene glycol, B: 1-bromonaphthalene, D: dimethyl sulfoxide

### 3.4. Polynomial fit of $\gamma_{SV}$ as a function of $\gamma_{LV}$ and $\theta$

The general form of the polynomial function $PM\gamma_{SV}(\theta, \gamma_{LV})$ to directly obtain the value of the surface free energy from $\theta$ and $\gamma_{LV}$ of a particular probe liquid adjusted from the third Neumann equation of state is written as follows by equation 15:

$$PM\gamma_{SV}(\theta, \gamma_{LV}) = A(\gamma_{LV}) \times \theta^5 + B(\gamma_{LV}) \times \theta^4 + C(\gamma_{LV}) \times \theta^3 + D(\gamma_{LV}) \times \theta^2 + E(\gamma_{LV}) \times \theta^1 + F(\gamma_{LV}) \qquad (15)$$

The coefficients of this polynomial function are not constants but are instead functions of $\gamma_{LV}$, which allows generalizing the polynomial function $PM\gamma_{SV}(\theta, \gamma_{LV})$ to be used with different probe liquids. The following polynomial fits were obtained for the coefficients A, B, C, D, E and F:

$$A(\gamma_{LV}) = -3.6735 \times 10^{-14}\gamma_{LV}^3 + 5.4452 \times 10^{-12}\gamma_{LV}^2 - 1.9749 \times 10^{-10}\gamma_{LV} +$$
$$1.5627 \times 10^{-9} \qquad (27)$$

$$B(\gamma_{LV}) = +1.1040 \times 10^{-11}\gamma_{LV}^3 - 1.6918 \times 10^{-9}\gamma_{LV}^2 + 6.1290 \times 10^{-8}\gamma_{LV} -$$
$$5.3196 \times 10^{-7} \qquad (28)$$

$$C(\gamma_{LV}) = -1.0814 \times 10^{-9}\gamma_{LV}^3 + 1.6463 \times 10^{-7}\gamma_{LV}^2 - 5.0646 \times 10^{-6}\gamma_{LV} +$$
$$5.6396 \times 10^{-5} \qquad (29)$$

$$D(\gamma_{LV}) = +4.2138 \times 10^{-8}\gamma_{LV}^3 - 5.6112 \times 10^{-6}\gamma_{LV}^2 + 2.8255 \times 10^{-5}\gamma_{LV} -$$
$$2.1287 \times 10^{-3} \qquad (30)$$

$$E(\gamma_{LV}) = -5.7967 \times 10^{-7}\gamma_{LV}^3 + 7.4232 \times 10^{-5}\gamma_{LV}^2 - 2.5385 \times 10^{-3}\gamma_{LV} +$$
$$3.1449 \times 10^{-2} \qquad (31)$$

$$F(\gamma_{LV}) = +9.9952 \times 10^{-1}\gamma_{LV} - 9.3025 \times 10^{-3} \qquad (32)$$

where $\theta$ is expressed in degrees ($deg$) and $\gamma_{LV}$ in $mJ/m^2$. For $A(\gamma_{LV}), B(\gamma_{LV}), C(\gamma_{LV}), D(\gamma_{LV}), E(\gamma_{LV})$ and $F(\gamma_{LV})$, the units of the independent term are $mJ/m^2$, while the units of the coefficients that multiply $\gamma_{LV}^3$, $\gamma_{LV}^2$ and $\gamma_{LV}$ are $m^2/(mJ^2\ deg^n), m/(mJ\ deg^n)$ and $1/deg^n$, respectively, where $n$ equals 5, 4, 3, 2, 1 and 0 for A, B, C, D, E and F respectively. The domain of the general polynomial is $1° \leq \theta \leq 130°$ and $25\ mJ/m^2 \leq \gamma_{LV} \leq 72.8\ mJ/m^2$.

Figure 10 shows the relative discrepancy $RD^{\gamma_{SV}^{III}-PM\gamma_{SV}}$ between the values of $\gamma_{SV}^{III}$ and the values of $\gamma_{SV}$ obtained by means of the polynomial $PM\gamma_{SV}(\theta, \gamma_{LV})$. It can be seen that, in the area of experimental interest, where $\gamma_{SV} \geq 10\ mJ/m^2$, the relative discrepancy takes values between -0.5% and +0.5%. The relative discrepancy values lower than -2.0% and greater than +2.0% are found in the area where the values of $\gamma_{SV}$ are lower than or equal to $5\ mJ/m^2$. The relative discrepancy values between those for

$\gamma_{SV}$ obtained by the polynomial $PM\gamma_{SV}(\theta, \gamma_{LV})$ and those obtained by equation 9 ($\gamma_{SV}^{III}$) are generally smaller than the relative discrepancies between the different equations of state $RD^{III-I}$, $RD^{II-I}$ and $RD^{III-II}$ (see Figures 5, 6 and 7). This indicates that the general polynomial $PM\gamma_{SV}(\theta, \gamma_{LV})$ is a model very close to the most recent Neumann equation of state, $\gamma_{SV}^{III}$, even with less discrepancy than the previous equations of state.

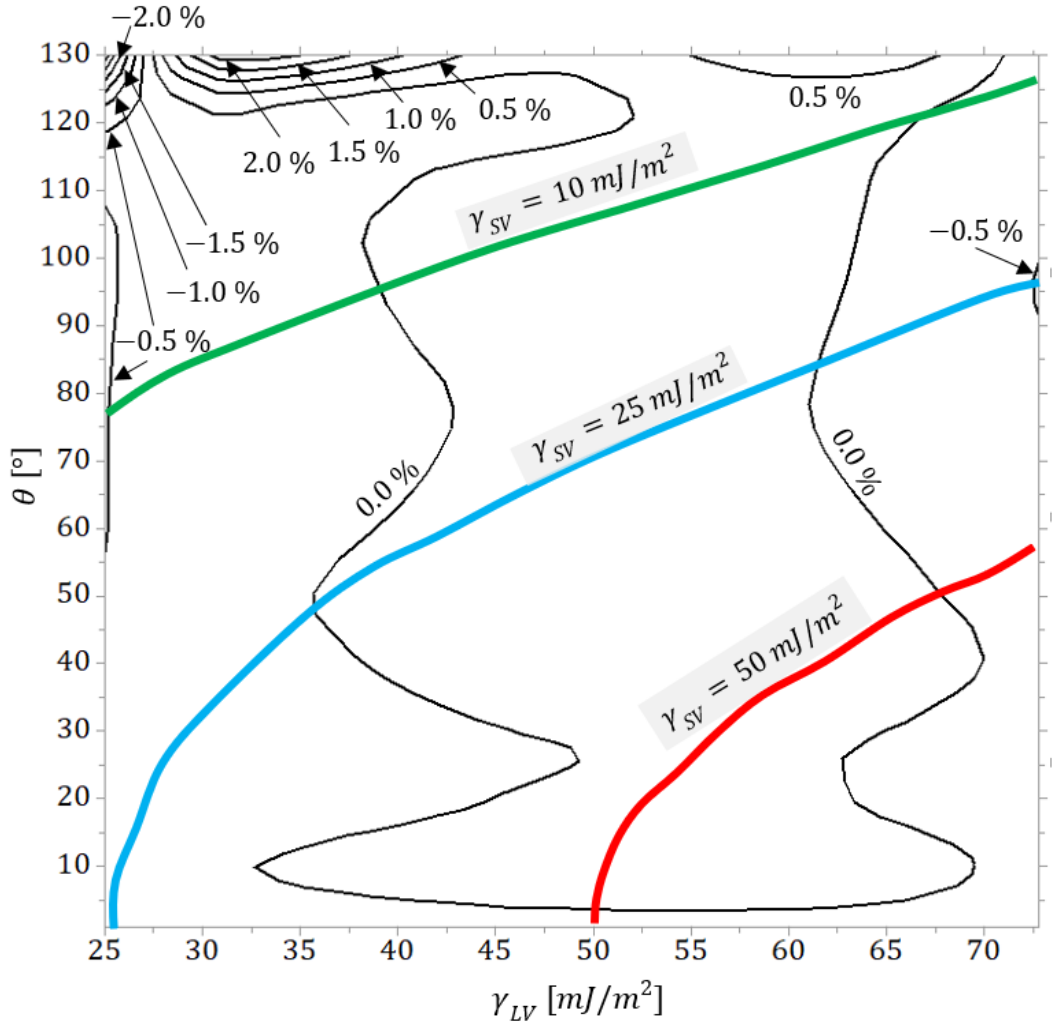

Figure 10. Level curve graph of the relative discrepancy $RD^{\gamma_{SV}^{III}-PM\gamma_{SV}}$ as a function of $\theta$ and $\gamma_{LV}$. The colored lines indicate a given value of $\gamma_{SV}$ (green: $\gamma_{SV}=10$ mJ/m², light blue: $\gamma_{SV}=25$ mJ/m², red: $\gamma_{SV}=50$ mJ/m²). The plot is constructed by Akima interpolation [53] from $11\times130=1430$ values of $RD^{\gamma_{SV}^{III}-PM\gamma_{SV}}$.

A histogram of the discrepancy values $D^{\gamma_{SV}^{III}-PM\gamma_{SV}}$ is shown in Figure 11. The maximum value is 0.03782 $mJ/m^2$, while the minimum is -0.03458 $mJ/m^2$. The standard deviation of the distribution of values is 0.01257 $mJ/m^2$, which can be considered as the standard uncertainty to obtain the values of $\gamma_{SV}$.

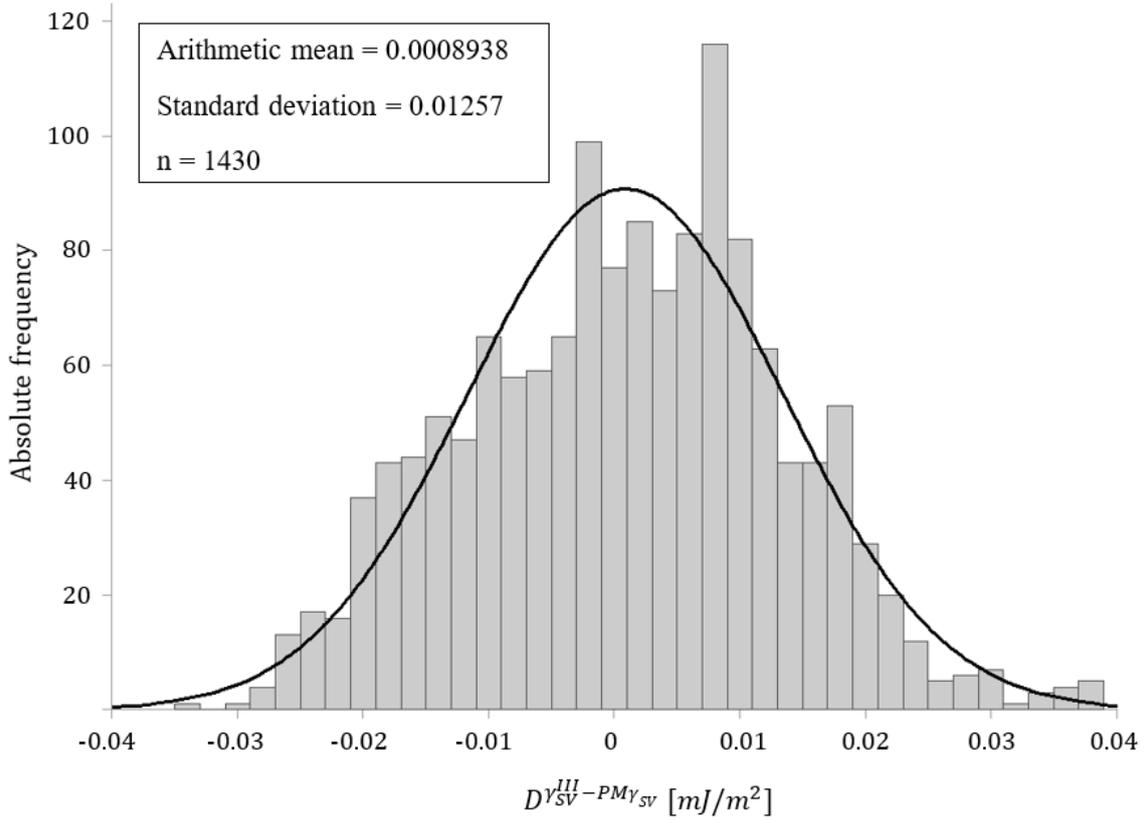

*Figure 11. Histogram of the discrepancy values (D) between the surface energy values obtained through the numerical resolution of equation 9 ($\gamma_{SV}^{III}$) and those obtained through the general polynomial ($PM\gamma_{SV}(\theta, \gamma_{LV})$). A solid black line shows a normal distribution with arithmetic mean and standard deviation equal to the data.*

In addition to the ease of use of the polynomial function $PM\gamma_{SV}(\theta, \gamma_{LV})$ with respect to the numerical resolution required when using Neumann's third equation of state (equation 9), the function $PM\gamma_{SV}(\theta, \gamma_{LV})$ allows quickly performing a simple propagation of the uncertainties of $\theta$ and of $\gamma_{LV}$ in the determination of $\gamma_{SV}$ by using the Taylor series method [57,58].

It is worth mentioning that, in addition to the polynomial fit $PM\gamma_{SV}(\theta, \gamma_{LV})$ presented, other polynomials were obtained using a surface fit approach (response surfaces) and applying the least squares method. Several polynomials as functions of $\theta$ and $\gamma_{LV}$ up to the fifth order were obtained. However, none of them gave sufficiently low relative discrepancy results compared to $\gamma_{SV}^{III}$. In general, the relative discrepancy was acceptable ($RD \cong \pm 1\%$) in areas where $\theta$ and $\gamma_{LV}$ took intermediate values, while it was high ($RD \cong \pm 10\%$) when $\theta$ and $\gamma_{LV}$ took higher and/or lower values.

**Conclusions**

The largest values of relative discrepancy between the different Neumann equations of state are observed at high contact angle values (greater than ~100º). The relative discrepancy between the different equations of state decreases appreciably with increasing values of $\gamma_{SV}$. In general, the relative discrepancy is larger between equations of state III and I, intermediate between equations of state II and I, and smaller between equations of state III and II.

The sensitivity analysis using the condition number of the three Neumann equations of state revealed the existence of lower bounds of uncertainty in the determination of $\gamma_{SV}$ due to the uncertainty in the experimental determination of the contact angle. For a certain surface ($\gamma_{LV}$ = constant), the uncertainty level depends on the probe liquid used and is lower for liquids with lower $\gamma_{LV}$.

The fifth-order polynomial fit provided the best results in terms of discrepancy and relative discrepancy vs. the equation of state III when compared with the polynomials of order 2, 3 and 4. This makes it preferable to generalize the procedure to other test liquids.

The fifth-order polynomial fits of the equation of state III for the seven most used probe liquids allowed the direct calculation of $\gamma_{SV}$ with very low relative discrepancy with respect to the numerical solution, with values of discrepancy not exceeding ±1% in any of the cases.

The general polynomial fit of the equation of state III allows calculating $\gamma_{SV}$ directly and for any particular probe liquid. Furthermore, it is a simple function to use and to propagate the error associated with $\gamma_{LV}$ and $\theta$. It is highlighted that the relative discrepancy with respect to the numerical resolution for the cases where $\gamma_{SV}$ is greater than $10\ mJ/m^2$ takes values of ±0.5%, while the absolute discrepancy is ± 0.04 $mJ/m^2$. Finally, we conclude that both the polynomial approximations for particular probe liquids and the general approximation can be easily derived or integrated for analytical application both for the calculation of point values and for use in mathematical models.

**Appendix A. Analysis of the equation of state I**

To observe the global behavior of equation 5, it is identified that the rational term generates a discontinuity, which can be more easily analyzed graphically from the following equation:

$$f_I(\gamma_{LV}, \gamma_{SV}) = \frac{(0.015\gamma_{SV} - 2.00)\sqrt{\gamma_{LV}\gamma_{SV}} + \gamma_{LV}}{\gamma_{LV}(0.015\sqrt{\gamma_{LV}\gamma_{SV}} - 1)} \qquad (A.1)$$

This equation can be plotted in the experimental region of interest (i.e. $1\ mJ/m^2 < \gamma_{SV} < 73\ mJ/m^2$ and $25\ mJ/m^2 < \gamma_{LV} < 73\ mJ/m^2$).

The graph of equation A.1 is presented in Figure 1.A, which shows the presence of a discontinuity on the surface, which is the product of the fact that equation 5 is indeterminate when the denominator is zero. This happens when $\gamma_{SV} = \frac{1}{0.000225}\gamma_{LV}$. As can be seen in Figure 1.B, equation 5 ceases to be useful not only where this condition is met, but also in the vicinity. Neuman et al. [14] proposed to overcome this problem by obtaining physically correct curves by interpolation between regions sufficiently far from the discontinuity; however, performing such interpolation is not trivial. An alternative solution is to use the tables published by Neumann et al. [45] where the values of $\gamma_{SV}$ have already been calculated for $\theta$ from 10º to 110º, with liquids whose $\gamma_{LV}$ ranges from 30 $mJ/m^2$ to 73 $mJ/m^2$.

In the same way and for comparison, equation 9 can be analyzed without the term $\cos\theta$ so that it is equivalent to equation A.1.:

$$f_{III}(\gamma_{LV},\gamma_{SV}) = 2\sqrt{\gamma_{SV}/\gamma_{LV}}\,(1-\beta_2(\gamma_{LV}-\gamma_{SV})^2) - 1 \qquad (A.2)$$

The graph in the experimental zone of interest of equation A.2 is presented in Figure 1.B, where a smooth surface with no discontinuities can be observed. The behavior of equation 7 (not shown) is very similar to that of equation 9.

**Appendix B. Procedure to obtain the general polynomial $PM\gamma_{SV}(\theta,\gamma_{LV})$**

Figure B.1 schematically shows the procedure to obtain the general polynomial $PM\gamma_{SV}(\theta,\gamma_{LV})$.

I-) The 11 discrete functions $\gamma_{SV}(\theta)$, one for each hypothetical probe liquid and obtained by numerical resolution of equation 9 ($\gamma_{SV}^{III}$) for $\theta$ values ranging from 1º to 130º, were fitted using a polynomial of fifth order.

II-) Thus, 11 polynomial functions $P_5\gamma_{SV}(\theta)$ were obtained. The analysis of these equations allows considering the search for the functional relationship between their coefficients and the variation in $\gamma_{LV}$. This leads to the assumption of a general polynomial of two variables $PM\gamma_{SV}(\theta,\gamma_{LV})$.

III-) A polynomial adjustment of the coefficients vs. $\gamma_{LV}$ is used to find the functions to describe the change in the coefficients of the polynomials $P_5\gamma_{SV}(\theta)$ as a function of $\gamma_{LV}$.

IV-) Six polynomial functions that describe the relationship between the coefficients and $\gamma_{LV}$ were obtained.

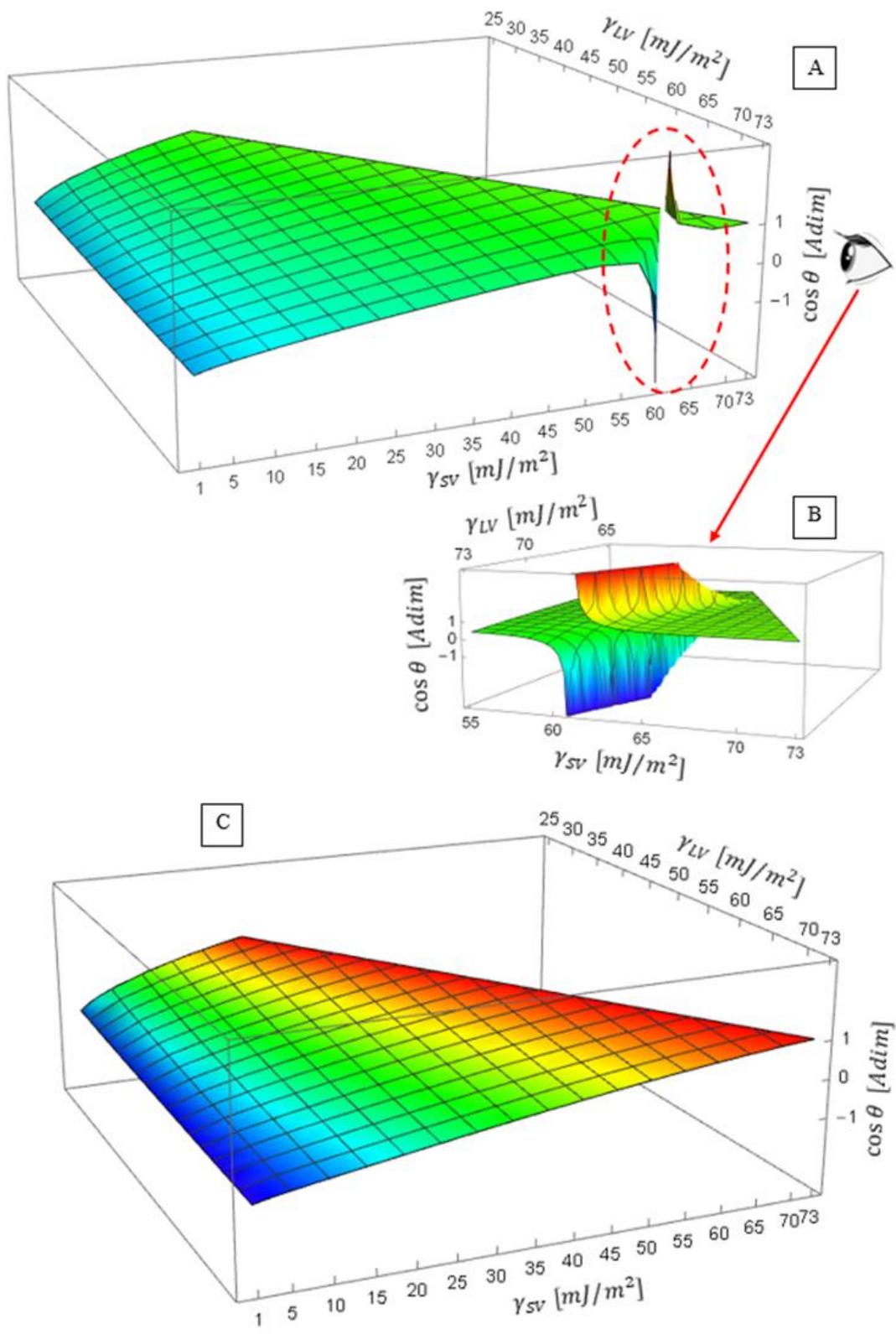

Figure 1. A-) Graph of $f_I(\gamma_{LV}, \gamma_{SV})$. B-) Detail of the discontinuity of $f_I(\gamma_{LV}, \gamma_{SV})$. C-) Graph of $f_{III}(\gamma_{LV}, \gamma_{SV})$.

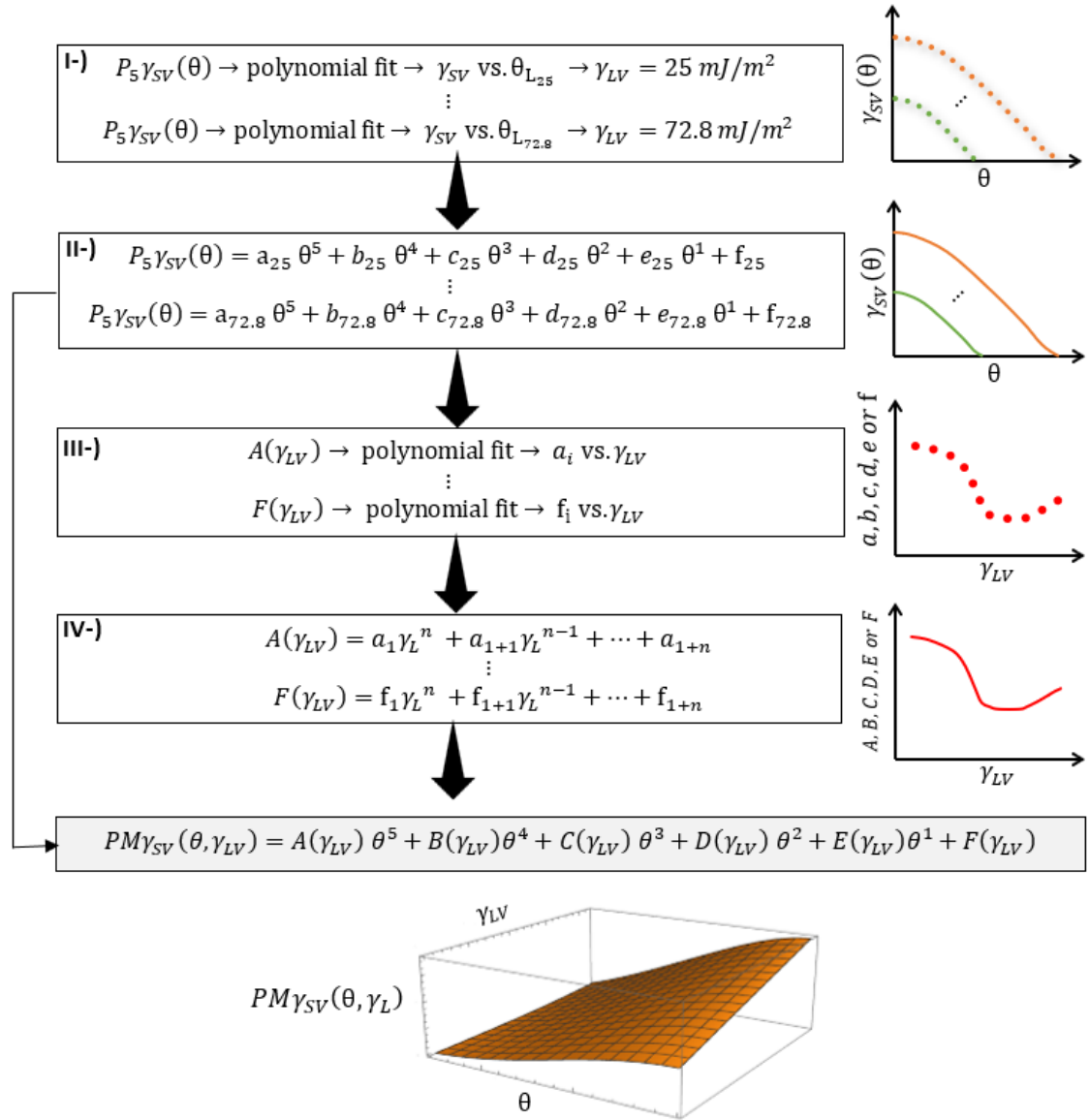

*Figure B.1. General scheme to obtain the polynomial function $PM\gamma_{SV}(\theta, \gamma_{LV})$.*

**Acknowledgements**

The authors would like to thank CONICET and Agencia Nacional de Promoción Científica y Tecnológica (ANPCyT) from Argentina. This work has been funded by Agencia Nacional de Promoción Científica y Tecnológica (ANPCyT) from Argentina through PICT-2017-2494 and PUE-CONICET-2019-574-APN projects.